\documentstyle[prd,aps,preprint]{revtex}
\tightenlines
\input epsf.tex
\def\DESepsf(#1 width #2){\epsfxsize=#2 \epsfbox{#1}}
\begin{document}

\draft
%@twocolumnfalse\endcsname
\preprint{\vbox{
\hbox{OSU-HEP-98-10}\hbox{CTP-TAMU-50-98}
\hbox{UMD-PP-99-057}
}}
\title{Partial Yukawa unification
and a supersymmetric origin \\ of flavor mixing}
\author{ K. S. Babu$^1$, B. Dutta$^2$ and R. N. Mohapatra$^3$ }

\address{$^1$Department of Physics, Oklahoma State University, Stillwater, OK, 74078\\
$^2$Center for Theoretical Physics, Department of Physics, Texas A \& M
University, College Station, TX, 77843\\
$^3$Department of
Physics, University of Maryland, College Park, MD, 20742.}
\date{December, 1998}
\maketitle
\begin{abstract}
In a large class of supersymmetric SO(10) and left-right models, requiring
that the effective theory below the scale of $SU(2)_R$ breaking be the MSSM
implies partial Yukawa unification with $Y_u=Y_d$ and $Y_e=Y_{\nu^D}$. The
same result also emerges in models with a horizontal $SU(3)_H$ symmetry.
As a result, at the tree level, these models lead to vanishing quark mixing
angles. We show that the correct mixing pattern can be generated in
these models at the loop level from the flavor structure associated with the
supersymmetry breaking terms.  We generalize the constraints on supersymmetric
parameters from flavor changing neutral current (FCNC) processes to include
several squark mass insertions and confirm the consistency of the scheme.
The expectations of this scheme for CP violating observables in the $B$
meson system are quiet different from the KM model, so it can be tested at
the B factories.
\end{abstract}

\vskip2pc

\section{Introduction}

One of the fundamental problems of the standard electroweak model
is the lack of understanding of the fermion mass hierarchies and the flavor
mixings. The minimal supersymmetric standard model (MSSM) which provides
interesting resolution of the Higgs and
the symmetry breaking puzzles does not shed any light on
this question \cite{martin}. Supersymmetric grand unified theories do provide
one framework within which question of flavor has been addressed in a variety
of ways. While there are several interesting ideas, no particularly compelling
model has emerged so far. It is worthwhile to pursue alternative approaches to
the flavor problem.  In this paper, we discuss one such approach.

We discuss two extensions  of the MSSM which may provide an
interesting new solution to the flavor question.  At the tree level
they lead to the equality of the up and the down type Yukawa matrices
which we call up-down unification. While this class of models does not
explain the microscopic origin of fermion mass hierarchies, the up-down unification
implies that the quark mixing angles and the CP violating phase vanish at the tree
level and arise purely as a result of radiative corrections.  This would explain
the smallness of these mixing parameters naturally.  Flavor mixing associated with
the soft supersymmetry breaking terms play a crucial role in inducing such mixings;
we will show that they are compatible with existing flavor changing neutral current
constraints (FCNC).  We will argue that the required
flavor structure arise quite naturally in a variety of supersymmetry
breaking schemes.

One class of models which exhibits up-down unification is the supersymmetric left--right
model with the seesaw
mechanism\cite{gell} for small neutrino masses
that has recently been discussed as a possible solution to
the question of R-parity violation\cite{moh} and strong and weak CP
violation problems of the MSSM\cite{rasin}. The second class of models
uses a horizontal $SU(3)_H$ symmetry (local or global). The latter belongs to the
class of models which generically leads to relations between the masses of
different generations of quarks. In our discussion however, we do not make the
additional assumptions that lead to such mass relations, for example,
specific pattern of VEVs for the Higgs fields.  Rather, after we choose
the Higgs sector, we allow them to have arbitrary VEVs and follow its
consequences. It turns out that they then automatically lead to up-down
unification in the sense we are going to discuss.

The main result that we wish to exploit in providing a radiative origin
of quark mixings is the following. If we assume that below the new theory
scale, i.e., the $SU(2)_R$ scale $v_R$ in the case of SUSYLR models and
$SU(3)_H$ scale $v_H$ in the case of horizontal symmetry model, the theory
is given by MSSM, then it is easy to show that there is up-down
unification that leads to:
\begin{eqnarray}
Y_u&=& Y_d\nonumber\\
Y_l&=& Y_{\nu^D}~.
\end{eqnarray}
Strictly speaking, for the $SU(3)_H$ case,
$Y_u = \beta_q Y_d$ and $Y_l=\beta_lY_{\nu^D}$, where $\beta_{q,l}$ are
ratios of two Yukawa couplings.  But by redefining $\tan\beta \equiv \left\langle
H_u \right \rangle/\left\langle H_d \right \rangle$ one can set $\beta_q=\beta_l=1$

While the up down unification is obvious in $SO(10)$ with a single {\bf 10} of Higgs representation,
its generality when any number of {\bf 10}'s
and {\bf 126} Higgses are included was noted in Ref. \cite{lee}.  This point was made in
the context of SUSYLR model in Ref.\cite{goran2}. To the best of our knowledge, the
$SU(3)_H$ example has not been discussed before.

Eq. (1) implies that at the tree
level, the quark mixing angles all vanish. We
then show that consistent with the known constraints from flavor
changing neutral currents and vacuum being charge and color preserving and stable,
it is possible to generate desired values for the quark mixings.
In this process we generalize  existing constraints on the soft SUSY
breaking terms from FCNC effects by allowing for several squark mass insertions
and find that some may be relaxed.

This paper is organized as follows: in Sec. 2, we present the arguments
leading to Eq. (1) in the supersymmetric left-right and the SO(10) models;
in Sec. 3, we derive the same relations for the $SU(3)_H$ model. In Sec. 4,
we outline our choice of soft supersymmetry breaking terms; in Sec. 5, we
discuss the one loop graphs that induce the correct flavor mixings. In
Sec. 6 we discuss the one loop effects from the off diagonal squark masses
and the constraints on their magnitude from flavor changing neutral current
effects; in Sec. 7, the same discussion is repeated for the off diagonal
A-terms; and in Sec. 8 we discuss the origin of CP violation.

\section{Deriving Up-Down unification in the left-right and SO(10) models}

In this section, we present a brief review of the arguments leading to
the up-down unification in the SUSYLR model followed by its extension to
the SO(10) model. We start by writing the superpotential of the SUSYLR model
based on the gauge group$SU(2)_L\times SU(2)_R\times U(1)_{B-L}\times
SU(3)_c$ (we have suppressed the generation index):

\begin{eqnarray}
W & = &
{\bf h}^{(i)}_q Q^T \tau_2 \Phi_i \tau_2 Q^c +
{\bf h}^{(i)}_l L^T \tau_2 \Phi_i \tau_2 L^c
\nonumber\\
  & +  & i ( {\bf f} L^T \tau_2 \Delta L + {\bf f}_c
{L^c}^T \tau_2 \Delta^c L^c)
\nonumber\\
  & +  & M_{\Delta} {\rm Tr} ( \Delta \bar{\Delta} ) +%\ {\rm Tr} (
\Delta^c \bar{{\Delta}^c }) +\lambda S(\Delta\overline{\Delta}
-\Delta^c\overline{\Delta^c}) + \mu_S S^2 +
\mu_{ij} {\rm Tr} ( \tau_2 \Phi^T_i \tau_2 \Phi_j )
\nonumber\\
 & + & W_{\it NR}
\label{eq:superpot}~.
\end{eqnarray}
Here we use the standard notation (see Ref. \cite{kuchi}) where $Q,Q^c$ denote the left
handed and right handed quark doublets, $L, L^c$ denote the lepton doublets,
$\Phi_i$ are the (2,2,0,1) Higgs bi-doublets,
$\Delta$ and $\Delta^c$ are the left and right handed Higgs triplets \{(3,1,2,1) + (1,3,-2,1)\}
and $\bar{\Delta}$ and $\bar{\Delta^c}$ are fields conjugate to $\Delta, \Delta^c$.
$W_{\it NR}$ denotes non-renormalizable terms arising from
higher scale physics such as grand unified theories or Planck scale effects.
The need for the nonrenormalizable term has been discussed in
Ref.\cite{kuchi} and we do not repeat it here since it plays no role in our
discussion.

We will work in the vacuum which conserves R-parity which implies that
$v_R\geq 10^{10}$ GeV. In order to demonstrate the Yukawa unification,
let us discuss the Higgs doublet spectrum of the model at low energies.
Suppose that at the scale $v_R$, where $SU(2)_R$ breaks,
we have  an arbitrary number of bi-doublet
fields $\Phi$'s. In order to get the MSSM at low energies, one must decouple
all but one pair of $H_{u}$ and $H_d$ from the low energy spectrum. This has
been called doublet-doublet splitting problem in literature. It is clear
from the superpotential in Eq. (2) that doublet Higgsino matrix is symmetric
because of parity invariance. For two bi-doublets for instance,
it looks like:
\begin{eqnarray}
M_H= \left(\begin{array}{cc}
\mu_{11} & \mu_{12}\\
\mu_{12} & \mu_{22} \end{array}\right)~.
\end{eqnarray}
Eq. (3) is the mass matrix of bi-doublet $Higgsinos$, but since supersymmetry
is broken at a much lower scale, the Higgs bosons will be degenerate with the
Higgsinos.  If we now do fine tuning to get one pair of $H_{u,d}$ at low energies,
the $H_{u,d}$ appear as identical combinations of the doublets in $\Phi_i$'s.
(Recall that a complex symmetric matrix may be diagonalized by  the transformation
$U^T M_H U = M_H^{\rm diag}$.  This means that the rotations on $\Phi_1^u$ and
$\Phi_2^u$ to get the MSSM $H_u$ is identical to the rotation on $\Phi_1^d$
and $\Phi_2^d$ to get $H_d$.)
As a result, at the MSSM level, we have up down Yukawa unification, Eq.
(1).  It is clear that this result holds in the
presence of arbitrary number of bi-doublet fields.

A similar situation appears in the case of a class of SUSY SO(10) models.
Let us assume that the Higgs sector of the model has only {\bf 10} and
$\overline{\bf 126}$'s and if there are {\bf 16}-dimensional spinor fields, they do not participate
in the low energy doublets (i.e., they do not mix with the {\bf 10} or $\overline
{\bf 126}$).
The main point here is that above the GUT scale there may be more than
one pair of MSSM doublets contained in some {\bf 10} or
in one of the $\overline{\bf 126}$. In order to get the MSSM below
the GUT scale, one will have to diagonalize the doublet mass matrix so that
one pair of light doublets remains light (doublet-doublet
splitting). This may need a fine tuning of
parameters, or it may appear naturally via the Dimopoulos-Wilczek mechanism \cite{dw}.
We will not concern ourselves with its detail. {\it Regardless of how
many pairs of doublets couple to fermions at the GUT scale, as long as the doublet mass
matrix is symmetric, the effective Yukawa couplings of the up and
down sector Higgs ($H_u$ and $H_d$) of the MSSM are the same} \cite{lee}.
This leads to the afore-mentioned up-down
unification\cite{others}. Note that if the {\bf 16} and $\bar{\bf 16}$ fields
participate in the low energy doublets, the $H_u$ and $H_d$ couplings are in
general unrelated to each other\cite{babu}.

There is one other class of SO(10) models where up-down unification of Eq. (1) would hold.
Suppose there is a single {\bf 10} of Higgs that couples to fermions.  In this case,
even if the {\bf 16} spinors participates in electroweak symmetry breaking (by
mixing with the {\bf 10}), because there is only one Yukawa coupling matrix to
begin with, Eq. (1) would hold.  Note that in this case $\tan\beta$ is not restricted
to be equal to $m_t^0/m_b^0$, this situation is analogous to the case of $SU(3)_H$
(see below).

\section{Horizontal symmetry and up-down unification}

Let us now present a horizontal $SU(3)_H$ (local or global) model
that leads to up-down unification. Consider MSSM$\otimes SU(3)_H$
as the high scale theory. Parity invariance is not necessary here.
We assume that the matter multiplets of the MSSM transform as the
{\bf 3} dimensional representation under the group \cite{bere}. Let us assume
that we have Higgs doublet fields transforming as $\Phi_{u,d}
(2, \pm 1, 6^*)$ under the gauge group $SU(2)_L\times U(1)_Y\times
SU(3)_H$. Let us include an isosinglet field pair $S(1, 1, 6^*)$
and ${\bar S}(1, 1, 6)$. The Horizontal symmetry is broken down at
some high scale by the VEVs $\left\langle S \right\rangle
=\left\langle \bar{S}\right \rangle= v_H X$
where $X$ is a symmetric $3\times 3$ matrix that does not commute
with any of the SU(3) generators. The superpotential for the
theory contains the following terms:
\begin{eqnarray}
W_H = h_u Q \Phi_u u^c + h_d Q \Phi_d d^c + \Phi_u \Phi_d S + T(S\bar{S}
-v^2_H)~,
\end{eqnarray}
where $T$ is a singlet used to break the $SU(3)_H$. In general one needs
other fields such as $H-$octets etc to get the correct symmetry breaking
pattern. There is no need for us to be explicit about them. It is clear
from the superpotential that once $S$ field acquires arbitrary  VEVs, we will
get a symmetric $6\times 6$ mass matrix for the $\Phi_{u,d}$ fields.
Since we
want the MSSM to emerge at low energies, there will be some kind of fine
tuning necessary at the $SU(3)_H$ scale. Due to the
symmetry of the Higgs mass matrix, the MSSM Higgses $H_{u,d}$
will come out as the identical linear combination of the various
$\Phi_{u,d}$
fields. This will then lead to the up-down unification Eq. (1) with
$\beta_q=\frac{h_d}{h_u}$. A similar equation will also follow for the
lepton sector.  Even if there are an arbitrary number of $S$ fields, this
result would hold.  Note that $\tan\beta$ is a free parameter here,
not restricted to be equal to $m_t^0/m_b^0$.

\section{Nature of supersymmetry breaking terms}

An immediate consequence of the above result (Eq. (1))
is that at the tree level the
quark mixing angles vanish. However, once soft supersymmetry breaking
terms are taken into account, they lead to correction to the tree level
predictions via one loop diagrams. These diagrams
not only correct the tree level mass spectrum bringing it more into
agreement with observations but they also provide an explanation of the
smallness of the CKM angles in a natural manner. In doing this,
we will invoke  two types of supersymmetry
breaking terms: (i) The bilinear mass terms involving the superpartner
fields, and (ii) the trilinear terms
(the $A$ terms) that arise in generic supergravity and string models.

It would be necessary to have the generation changing bilinear terms to be
much smaller than the generation diagonal bilinears in order to be compatible
with known FCNC constraints.  This type of a spectrum can be achieved
in different scenarios of supersymmetry breaking: e.g. (1) dilaton dominated
supersymmetry breaking \cite{casas1},
(2) anomalous $U(1)$ model\cite{riotto}, (3) models with
nonabelian flavor symmetry\cite{hall}, (4) gauge and gravity mediated
supersymmetry breaking \cite{jmr}. In all these models generational
mixing bilinear terms are much suppressed (by a loop factor  in (1),
a chirality factor in (3) and a ratio of scales in (2) and (4)).
The Kahler potential for these types of models can be written as:
 \begin{eqnarray}
K(z,z^*; y_i, y^*_i)= M^{-2}_{Pl}(z^*z +y^*_iy_i + \epsilon_{ij}y^*_iy_j)
\end{eqnarray}
where $\epsilon_{ij}$ are elements of
a Hermiitian matrix, $y_i$ stand for the visible
sector fields and $z$ is the hidden sector field.  The $\epsilon_{ij}$
contributes to departures from universality for the soft scalar masses.
The scalar masses can be written as follows:
\begin{eqnarray}
m^2_{ij}=m^2_0(\delta_{ij}+\epsilon_{ij})~
\end{eqnarray}
with $\epsilon_{ij} \ll \delta_{ij}$.

If we use the trilinear $A$ terms for the generation of the quark (and lepton)
mixing angles, the important requirement is that  the generation
dependence of the $A$ terms be different from the effective Yukawa
couplings below the scale
$v_R$. A significant point is that in the presence of a general Kahler
potential there is no reason for the effective supersymmetry breaking
mass or the $A$ terms surviving at the MSSM level to be proportional
to the Yukawa couplings since the unitary matrices that diagonalize the
Higgs mass matrix will not in general diagonalize the $A$ matrices. One
can therefore write
 \begin{eqnarray}
A^{(a)}_{p,ij}=m_1Y^{(a)}_{p,ij}+m_1\epsilon_{ik}Y^{(a)}_{p,kj}~.
\end{eqnarray}
Here $m_1$ is a supersymmetry breaking mass parameter, the subscript $p$ stands
for $u,d,l$ (quarks or leptons), and the superscript $(a)$ denotes the
Yukawa coupling to the Higgs field $\Phi_a$.
Below the $v_{R, H}$ scale, the diagonalization of the bilinear $\mu$
terms to
get one pair of $H_u,H_d$ diagonalizes only the first term in the above
expression for $A$ and leaves the second term as arbitrary. Anticipating that
FCNC effects require $\epsilon \ll 1$, we will then expect departures
from the proportionality between $Y$ and $A$ to be of order $\epsilon$
without any extra strong suppressions due to $m_q/m_W$. The main reason for this
is that we only expect one linear combination of the $v_R$ scale
$Y_p$'s to lead to quark masses and the orthogonal one remains free. The
$A$ term gets contributions from both the linear combination of $Y_p$'s.
For example, in the SUSYLR model (Eq. (2)), suppose there are two bi-doublets $\Phi_i$.
The MSSM doublet $H_d$ will be a linear combination $H_d = \cos\theta \Phi_1^d
+ \sin\theta \Phi_2^d$ (and similarly for $H_u$).
If the angle $\theta$ is small, the Yukawa couplings
$h^{(2)}_{q,ij}$ can be large, not proportional to the small quark masses.
There could be other ways of generating departure from proportionality
between the $A$ terms and the Yukawa couplings such as through the
superpotentials as in models where the supersymmetry breaking is mediated by
anomalous $U(1)$ gauge symmetries using more than one singlet.

In principle, both sources of flavor mixing will be present (Eqs, (6) and (7)).
But for conceptual simplicity, we shall assume one among the two to be dominant.
The results will change very little if both are combined.  Furthermore, for
most part, Eq. (6) and Eq. (7) lead to the same phenomenology as far as FCNC
processes are concerned.  One difference is in their cosmological implications.
Color and charge breaking does restrict the nature of the trilinear $A$
terms, but not the bilinear terms, we shall discuss these in  Sec. VII.
%\An important requirement on the $A$-terms is that their generation
%\dependence be different from the effective Yukawa couplings below the scale
%\$v_R$. We carefully examine the impact of choice of the $A_{ij}$ needed to
%\generate the correct quark mixing pattern on the flavor changing neutral
%\currents\cite{masiero} and the color and charge breaking by
%\vacuum\cite{casas}. We find that it is necessary for the theory to stay
%\in the metastable color conserving vacuum. We however have checked by
%\rough calculations that tunneling to the lower color breaking vacuum
%\for our choice of parameters is takes longer than the age of the universe.
%\The model is therefore cosmologically viable.

Although we begin with a parity--invariant theory in the cases of SUSYLR and
SO(10) models, since $v_R$ will be assumed to be much larger than the SUSY
breaking scale, the soft SUSY breaking parameters need not respect parity.
Parity violation can show up in the soft breaking sector if they
involve the VEV $v_R$ that breaks $SU(2)_R$.  We will also discuss
the possibility of
maintaining parity invariance in the soft terms.

In the horizontal $SU(3)_H$ models, it is very natural to have the leading
(renormalizable) term in the Kahler potential to be $SU(3)_H$ flavor symmetric.
That would neatly resolve the SUSY flavor problem of supergravity
models.  $SU(3)_H$--violating terms can appear in the Kahler potential as
higher dimensional terms involving standard model singlet fields
suppressed by the Planck scale eg: $K \supset (M_{Pl}^{-4})(Q^\dagger
Q)(S^\dagger S)$.  If the
VEVs of such $S$ fields are somewhat smaller than $M_{Pl}$, the smallness of
$\epsilon_{ij}$ will be naturally explained.

\section{One loop contributions to quark masses and mixings}

Let us restate the tree level mass relations predicted in these models.
For simplicity, we focus on the SUSYLR or SO(10) type models:
\begin{eqnarray}
M^0_u &=& \tan\beta M^0_d \nonumber\\
M^0_{\nu^D} &=& \tan\beta M^0_{\ell}~.
\end{eqnarray}
It follows that both the quark mass matrices can be diagonalized by
the same unitary matrix and hence the earlier assertion that
$V_{CKM}={\bf 1}$. The same comment holds for the lepton sector too.
Furthermore, since the Dirac masses for neutrinos are not known
the second mass sum-rule involving the leptonic sector can help
predict the neutrino masses. The mixing between different lepton generations
would then
arise from the see-saw formula which involves the Yukawa coupling of
the $\Delta^c$ with off-diagonal elements (see the couplings ${\bf f}_c$ of Eq. (2)).
That could potentially explain why the quark mixing angles are small (since they are
induced at loop level), while the leptonic mixings are large (they already exist at
the tree level).  We will not discuss the leptonic mixing angles
any further, we plan to return to this interesting question
in a future paper.  Here we will focus instead on the quark sector.

Before discussing quark mixings, let us first discuss the consistency
of observed quark masses with the tree level predictions given in Eq. (8).
Note that the above equation gives $\frac{m^0_s}{m^0_c}=\frac{m^0_b}{m^0_t}$
at the scale $v_R$.  The superscript $0$ on the masses signify that
they are the tree level masses evaluated at $\mu \sim M_{SUSY}$.
Since $Y_t=Y_b$, when renormalization group extrapolated
down to the weak scale, the relation is unchanged. Using the fact that
at $ \mu =M_{SUSY} \simeq m_t$, $m_b(\mu)\simeq 2.9 $ GeV, $m_t\simeq 175$ GeV,
$m_c(\mu)\simeq .67$ GeV
and $m_s(\mu)\simeq .081$ GeV, the above relation among the tree level masses
is seen to be off by almost a factor of eight. Clearly, large loop corrections
must be invoked to resolve this discrepancy. Luckily, it has already
been pointed out in the literature\cite{many,banks} that in supergravity models
(specially the ones with large $\tan\beta$), there are large radiative
corrections to the bottom, strange and down quark masses arising through
one--loop diagrams involving the exchange of gluino and the squarks (see
Fig. 1).  Analogous
corrections to $m_u, m_c$ and $m_t$ lack the $\tan\beta$ enhancement, so we
focus on the corrections to $m_d, m_s$ and $m_b$.  There are
two ways to resolve the above mass conundrum: (i) $b$--quark mass
receives little correction from loops.  In this case, $\tan\beta\simeq m_t^0/m_b^0
\sim 60$ as in the SUSYLR model.  Eq. (8) would then imply
$m^0_s(\mu)\simeq .011$ GeV.  The rest -- actually the bulk -- of the
strange quark mass (about $.07$ GeV) can arise from the one loop gluino graph
such as in Fig. 1.
(ii) Loop corrections suppress the $b$--quark mass (Fig. 1 with
all external quarks being $b$).  This can happen for
moderate values of $\tan\beta \simeq 10$.
It will be consistent with one version of SO(10) model
that was discussed (with one {\bf 10} of Higgs which mixes with {\bf 16}), as well
as the $SU(3)_H$ example.  In this case the bulk of the strange
quark mass comes from the tree level.  $b$--quark mass receives large negative
loop correction.
There are two sources for $m_b$: the gluino--squark exchange and the chargino-stop
exchange.
\begin{eqnarray}
\delta m_b\simeq
\frac{2\alpha_s}{3\pi}\frac{m_{\tilde{g}}}{m^2_{\tilde{q}}}(m_b^0\mu \tan\beta
+A_{33}^{(d)} m_0) + {\lambda_t^2 \over 16 \pi^2} {\mu \over
m^2_{\tilde{q}} } (m_b^0\mu + A^{(u)}_{33} m_0\tan\beta)~.
\end{eqnarray}
In writing Eq. (9) we have assumed $m_{\tilde{q}} \gg m_{\tilde{g}}, \mu$, but
in our numerical estimate we use the exact expression.
It might appear that the gluino exchange would lead to identical corrections
for $\delta m_b/m_b$ and $\delta m_s/m_s$, but the trilinear $A_{33}^{(u),(d)}$
may not eqaul $A^{(u),(d)}_{22}$.  Furthermore, the chargino contribution,
which is comparable to the gluino contribution,
is
absent for the strange quark mass.  We will discuss the numerical values of
$m_s$ and $m_b$ after taking the loop correction into account in the next section
after allowing for flavor mixing in the squark sector (which turns out to be
significant for $m_s$).

%\ The tree level $b$ quark mass in this case is too large ($\simeq
%\12$ GeV) and we must have about 70\% of it cancelled by the one loop
%\contribution. Let us continue for the rest of the paper with the second
%\alternative. First let us see if such a large cancellation can arise from
%\the one loop level for the $b$-quark mass.

%\The one loop contribution to the $b$-quark mass is given by a guino exchange
%\and a wino exchange graph. Assuming the first one dominates due to the large
%\color coupling, we find that
%\\begin{eqnarray}
%\\delta m_b\simeq
%\\frac{2\alpha_s}{3\pi}\frac{m_bm_{\tilde{G}}}{M^2_{\tilde{q}}}(\mu tan\beta
%\+A_{33})
%\\end{eqnarray}
%\For $\mu=600$ GeV, $M_{\tilde{G}}\simeq 1500$ GeV and $M_{\tilde{q}}\simeq
%\600$ GeV, $\tan\beta\sim 25$ (assuming $A_{33}\equiv 0$), we get $\delta m_b\simeq 10$ GeV, which is comparable to the tree
%\level value. We will therefore stay with this choice of the parameters in
%\the rest of the paper.

As far as the first generation quarks go, if we choose the Yukawa coupling
at the tree level such that $m_u\simeq .002~GeV$ (the right value) at $M_{SUSY}$, then
we would get a tiny tree level value for $m_d$ . However, as is well known, the
entire $m_d$ can arise in the process of generating flavor mixing angles.  That is,
$m_d \simeq \theta_C^2 m_s$ works quite well. We will use this mechanism here
to generate $m_d$.  (This will require existence of both $\overline{d}_L s_R$
and $\overline{s}_L d_R$ terms in the down quark mass matrix.  Only the
former is needed for inducing $\theta_C$, see comments below.)
Analogous expressions for $m_s$, viz., $m_s \simeq V_{cb}^2 m_b$,
is too small to explain the magnitude of $m_s$.

Let us now turn to the quark mixings.  It is easier to
induce off--diagonal elements in the down--quark mass matrix (rather than the
up--quark matrix) for two reasons:  (i) The one-loop diagram that would induce
off--diagonal mixings have a $\tan\beta$ enhancement factor in the down sector
but not in the up sector, and (ii) the charm and top quark masses are much larger
in magnitude compared to the strange and bottom masses respectively, so the
off--diagonal elements that are
necessary in the up--quark mass matrix to generate the CKM mixings are
larger (by about a factor of 10) compared to the ones needed in the down--quark
sector.  As we already noted,
there are two ways to generate the
off-diagonal terms in the quark mass matrices at the one loop level:
(i) the off-diagonal terms in
the bi-linear squark masses and (ii)  the off-diagonal $A$ terms.
In both cases we have to make sure that the
magnitudes of the $A_{ij}$'s or $\Delta m^2_{ij}$'s needed are not in
conflict with constraints of flavor changing neutral current effects
\cite{masiero}. We will
see that there are parameter spaces where $A_{ij}$'s or $\Delta m^2_{ij}$
produce the required quark mixing without violating the bounds from the
flavor changing neutral current.  We now turn to address this important issue.
%\However,
%\ the $A_{ij}$s are  highly constrained by the UFB and CCB. So the primary
%\mechanism that we will use to generate the quark mixings will be the
%\off-diagonal squark masses $\Delta m^2_{ij}$.

\section{Quark mixings from off-diagonal squark masses}

Here we will generalize the FCNC constraints arising from $K^0-\bar{K^0}$ system
\cite{masiero}
to allow for several squark mass insertions that would be appropriate for our
discussions.  We will find that some of the constraints get relaxed because
of the multiple insertions.

Let us first list the minimal number of mixing parameters among the generations we
need in order to generate the quark mixings. These are determined as follows:
The Cabibbo angle $V_{us}$ requires an effective entry $\overline{d}_L s_R$ in the
down quark mass matrix.  Similarly, $V_{cb}$ requires an entry $\overline{s}_L b_R$.
Rather than using direct left--right squark mixings of the same flavor structure, we shall
make use of the large $\tilde{b}_L - \tilde{b}_R$ mixing that is already present in
the model.  If in addition, we have
$\tilde d_L-\tilde b_L$,
$\tilde b_L-\tilde s_L$, and
$\tilde b_R-\tilde s_R$, then effectively we can induce both $\overline{d}_L s_R$ and
$\overline{s}_L b_R$ in the quark sector.  For example, $\overline{d}_L s_R$ can arise
via the chain $\tilde{d}_L -\tilde{b}_L \rightarrow \tilde{b}_L -\tilde{b}_R \rightarrow
\tilde{b}_R - \tilde{s}_R$.  Note that this set of parameters will also induce $V_{ub}$
via a $\overline{d}_L b_R$ quark mass term which arises from
$\tilde{d}_L- \tilde{b}_L \rightarrow \tilde{b}_L-\tilde{b}_R$.  We
will discuss this constraint in detail, but let us note that in the process of
inducing $V_{us}$ and $V_{cb}$,  a non--zero value of $V_{ub}$ will
already be induced given by $V_{ub} \simeq V_{us} V_{cb}$, which is roughly
of the right order.

With this minimal set of SUSY flavor mixing parameters, we will
generate adequate values of all CKM mixing angles.  However, since
the $\tilde{d}_R$ field does not appear in this minimal set, there
will not be any significant correction to the $d$--quark mass.
This deficiency will be removed when we later introduce an additional
flavor mixing parameter $\tilde{d}_R-\tilde{b}_R$.  We will show that
$m_d \simeq \theta_C^2 m_s$ will be induced in this case, without
conflicting with the FCNC bounds.  But at first we stick to the
minimal set without $\tilde{d}_R-\tilde{b}_R$ mixing.

We will use $K^0-\bar K^0$ to constrain the  mixing terms.
This constraint turns out to be the most stringent.  The only type
of flavor changing operators generated by the above mixing terms
are:
 \begin{eqnarray}
        Q_1 & = & \bar d^{\alpha}_L
\gamma_{\mu} s^{\alpha}_L \bar d^{\beta}_{L} \gamma_{\mu} s^{\beta}_
L\; ,
        \nonumber \\
        Q_2 & = & \bar d^{\alpha}_L  s^{\alpha}_R \bar d^{\beta}_L s^{\beta}_R\; ,
        \nonumber \\
        Q_3 & = & \bar d^{\alpha}_L  s^{\beta}_R \bar d^{\beta}_L s^{\alpha}_R\; .
\end{eqnarray}
The effective Hamiltonian involving these set of operators
 with different structure of the flavor changing bilinears is
 given in \cite{masiero}. We write below the Hamiltonian, but after generalizing
the loop functions to account for several squark mass insertions.

\begin{eqnarray}
        H&=&-\frac{\alpha_s^2}{216 m_{\tilde{q}}^2}\Biggl\{
      \left(\delta^d_{12}\right)^2_{LL}
        \left(  24\,Q_1\,x\,f_6(x) + 66\,Q_1\,\tilde{f}_6(x) \right)
        \nonumber \\
        &+& \left(\delta^d_{12}\right)^2_{LR}
        \left(  204\,Q_2\,x\,f_6(x) - 36\,Q_3\,x\,f_6(x) \right) \Biggr\}
\end{eqnarray}
where $x=m^2_{\tilde{g}}/m_{\tilde{q}}^2$, $m_{\tilde{q}}$
is the average squark mass,
$m_{\tilde{g}}$ is the gluino mass, $(\delta^d_{ij})_{LL,LR}$ represent the
product of mass insertions $\Delta m^2_{ij}/m^2_{\tilde{q}}$ that give
rise to a particular operator and the functions $f_6(x)$ and
$\tilde{f}_6(x)$ are given by :
\begin{eqnarray}
f_6(x)&=&{1\over{60 (x-1)^7}}(197+25 x-300 x^2+100 x^3-25 x^4+
3 x^5+60 (1+5 x) \ln x)~,\nonumber  \\
\tilde{f}_6(x)&=&-{1\over{30(x-1)^7}}(-6 -125 x+80 x^2+60 x^3-
10 x^4+ x^5-60 x (1+2 x)\ln x)~.
\end{eqnarray}
The diagrams for  $K^0-\bar K^0$ mixing are shown
in the Fig. 2.
In Fig. 2a we use the mixing  $\tilde d_L-\tilde b_L$, $\tilde b_L-\tilde s_L$
 in both the squark lines. In Fig. 2b we use the mixing
$\tilde d_L-\tilde b_L$, $\tilde b_L-\tilde b_R$ and $\tilde b_R-\tilde s_R$
 in both the squark lines.  We list the resulting constraints in Table 1
 on the $\Re(\delta_{12}) = \Re(\Delta m_{12}^{d^2}/m_{\tilde{q}}^2)$ for an
 average squark mass $m_{\tilde{q}} = 700~GeV$. We have used $m_K=498$ MeV,
$f_K=160$ MeV and $m_s(1 GeV)= 150$ MeV as input.
 \begin{center}
 \begin{tabular}{|c|c|c|}  \hline
 $x$ & $\sqrt{\left|\Re  \left(\delta^{d}_{12} \right)_{LL}^{2}\right|} $
 &
 $\sqrt{\left|\Re  \left(\delta^{d}_{12} \right)_{LR}^{2}\right|} $ \\
 \hline
0.1& $2.8 \times 10^{-2}$ &$7.5\times 10^{-3}$ \\
0.3&$4.7\times 10^{-2}$ &$8.0\times 10^{-3}$ \\
2.0&$1.51\times 10^{-1}$ &$1.2\times 10^{-2}$ \\
4.0&$1.45\times 10^{-1}$ &$1.5\times 10^{-2}$ \\
8.0&$1.71\times 10^{-1}$ &$2.0\times 10^{-2}$ \\
10&$1.85\times 10^{-1}$ &$2.2\times 10^{-2}$ \\
\hline\end{tabular}
 \end{center}
 Table 1: Upper limits on the product of squark mixings from $K^0-\bar{K^0}$
for an average squark mass
 $m_{\tilde{q}}=700\mbox{GeV}$ and for different values of
 $x=m_{\tilde{g}}^2/m_{\tilde{q}}^2$. For other values of $m_{\tilde{q}}$,
 the limits can be obtained by multiplying the ones in the table by
 $m_{\tilde{q}}(\mbox{GeV})/700$. $\Re$ stands for the real part.
 QCD corrections (not included in the Table)
 relax the bound in column 1 by about 30\% and
 tighten it for column 2 by a factor of 2.

Using Table 1 and expressing
$(\delta^d_{12})_{LL,LR}$ in terms of flavor changing mass insertions
$\delta_{ij}$'s (where
 $\delta_{ij}=\Delta m^2_{ij}/m^2_{\tilde q}$, $\Delta m^2_{i,j}$ is the
flavor mixing term),
 we
can put bound on the mixing term $\delta_{d_Lb_L}\delta_{b_Ls_L}$ from
column 1 and on the product
$\delta_{d_Lb_L}\delta_{b_Lb_R}\delta_{b_Rs_R}$ from column 2.
Note that $\delta_{b_Lb_R}$ can be obtained from our discussion of
the loop correction to the $b$-quark mass. It is of order
$\sim 1/10$ for $\tan\beta \simeq 25$
(since $\delta_{b_Lb_R}={{m_b \mu \tan\beta}\over {m_{\tilde
q}^2}}$ and if we choose
 $\mu=$600 GeV, $\tan\beta=25$, $m_{\tilde{q}} = 700~GeV$
 we get the above value). We can use this to
put a bound on the term
$\delta_{d_Lb_L}\delta_{b_Rs_R}$ from  Table 1.  It is important to note that the bounds
obtained in  Ref. \cite{masiero} are relaxed in our case due to the
particular flavor structure.
We also find that the QCD correction \cite{bagger} to $(LL)^2$ term increases the bound
by
$\sim 30\%$. For the $(LR)^2$, the upper bound tightens by a factor of 2.

The off diagonal
elements in the quark mass matrices are generated  at the one loop level
via gluino-squark exchange.  It is given by:
\begin{eqnarray} M_{23}= \frac{2\alpha_s}{3\pi}
m_{\tilde{g}}\delta_{s_Lb_L}\delta_{b_Lb_R}f1_M~,
\end{eqnarray}

\begin{eqnarray} M_{12}= \frac{2\alpha_s}{3\pi}
m_{\tilde{g}}\delta_{d_Lb_L}\delta_{b_Lb_R}
\delta_{s_Rb_R} f2_M~.
\end{eqnarray}
The relevant diagrams are shown in Fig. 1.  The loop functions that enter into
the mass terms are given by:
\begin{eqnarray}
f1_M={1\over {(1-x)^2}}(1-x+ x \ln x)~,
\end{eqnarray}
\begin{eqnarray}
f2_M={1\over {2(1-x)^3}}(1-x^2+2 x \ln x)~.
\end{eqnarray}
The functions are given in the limit where all the diagonal squark masses are equal.
Since the mixing is
small ($\sim 1/{10}$), using the same parameter space one
finds that  the functions with unequal masses differ from the above ones
by less than 4$\%$.

The mixing angles $V_{cb}$ and $V_{us}$ arise from the down quark mass matrix
elements $M_{23}$ and $M_{12}$ respectively.
To generate the correct mixing angle for the quarks, they must have
the values: $M_{23}\simeq V_{cb}m_b$ and
$M_{12}\simeq V_{us}m_s$.
Since the products of flavor changing transition entries for squarks
are already constrained by the known bounds from FCNC processes in Table 1,
we have to see if they allow us to generate the needed magnitudes of the
$M_{ij}$. The $M_{23}$ element
proves to be the most restrictive on the product of $\delta_{ij}$'s and
once that is satisfied $M_{12}$ is automatically satisfied. In Table 2,
we present the {\it lower limits} for the products of the $\delta_{ij}$'s
required to generate the corresponding $M_{ij}$'s. (We use
$V_{cb}=0.034$ and $V_{us}=0.22$.)  In writing Table 2, we
have assumed $m_{\tilde{q}} = 700~GeV$.

 \begin{center}
 \begin{tabular}{|c|c|c|}  \hline
 $x$ & $|\left(\delta^{d}_{M23} \right)_{LR}| $
 &
  $|\left(\delta^{d}_{M12} \right)_{LR}| $\\
  \hline
0.1&$1.8 \times 10^{-2}$&$6.2\times 10^{-3}$ \\
0.3&$1.2\times 10^{-2}$&$4.7\times 10^{-3}$ \\
2.0&$8.6\times 10^{-3}$&$4.4\times 10^{-3}$\\
4.0&$8.3\times 10^{-3}$&$4.9\times 10^{-3}$\\
8.0&$8.4\times 10^{-3}$&$5.7\times 10^{-3}$\\
10&$8.6\times 10^{-3}$&$6.2\times 10^{-3}$ \\
\hline\end{tabular}
 \end{center}

\noindent Table 2: The lower limits on the products of $\delta_{ij}$'s  required to generate
the desired quark mixing angles (for $m_{\tilde{q}} = 700~GeV$).  The lower limits
from column 2 should be compared with the upper limits from column 2 of Table 1 (divided
by a factor of 2 for QCD correction).  Compatibility is seen for $x \ge 0.3$.

We can now express
$ \left(\delta^{d}_{M12}
\right)_{LR}\equiv \delta_{d_Lb_L}\delta_{b_Lb_R}\delta_{b_Rs_R}$ and
 $ \left(\delta^{d}_{M23} \right)_{LR}\equiv
\delta_{s_Lb_L}\delta_{b_Lb_R}$.
 The upper limit on the first term can be read off from the second
column of Table 1.  Comparing with the lower limit from column 2 of Table 2,
we find that the bound is well satisfied for
$x>0.3$.  Like before if the limit on the $\delta_{s_Lb_L}$ is desired
one should multiply the numbers in Table 2, first column by a
factor ($\sim  10$) corresponding to $\tilde{b}_L-\tilde{b}_R$ mixing.
% But there is no upper limit from the $K^0-\bar K^0$ diagram.

There are two contributions to $V_{ub}$, one goes as the product $V_{us} V_{cb}$
and the other is $M_{13}/m_b$.  Here $M_{13}$ is the loop contribution to
the (1,3) entry of the down quark mass matrix arising from a diagram
analogous to Fig. 1b (replace
$s_L$ by $d_L$ in Fig. 1b).  The estimate of $M_{13}$ is obtained from Eq. (13) by the
replacement $\delta_{s_L b_L} \rightarrow \delta_{d_L b_L}$.  Using the relation $V_{ub} =
V_{us} V_{cb} + M_{13}/m_b$, setting its magnitude equal to $0.006$,
and allowing for the two contributions to have a relative negative sign, we find that
$|M_{13}| \le 0.045 ~GeV$.  For $m_{\tilde{q}} = 700 ~GeV$ and $x=0.1$, this
corresponds to the limit $\delta_{d_L b_L} \delta_{b_L b_R} \le 0.012$.  On
the other hand, from the required value of $V_{us}$, we have a lower limit
(for the same set of parameters) $\delta_{d_L b_L} \delta_{b_L b_R} \delta_{b_R s_R}
\ge 6.2 \times 10^{-3}$ (see Table 2).  Comparing the two constraints, we find
that $\delta_{b_R s_R} \ge 0.5$ .  Thus there is large mixing in the $\tilde{b}_R-
\tilde{s}_R$ sector, but that is consistent with all phenomenological constraints.

The process $b \rightarrow s \gamma$ does provide some useful limits.  There is a
diagram similar to Fig. 1b with $s_L$ replaced by $s_R$ and the squark line
emitting a photon.  From Ref. \cite{masiero}, $b \rightarrow s \gamma$
puts a limit on the product
$\delta_{b_R s_R} \delta_{b_L b_R} \le 1.3 \times 10^{-2} (m_{\tilde{q}}/500~GeV)^2$.
For a squark mass of 1 TeV and for $\delta_{b_R s_R} \simeq 0.5$,
we see that $\delta_{b_L b_R}$ should be less than about
$0.1$.   This would seem to prefer the scenario with moderate $\tan\beta$, although
the large $\tan\beta$ scenario is not inconsistent if $\mu$ is much smaller than
the squark masses.

As for the diagonal strange quark mass, it arises from a diagram analogous to
Fig. 1b.  The magnitude is given by Eq. (14) with $\delta_{d_Lb_L}$
replaced by $\delta_{s_L b_L}$.  We will now show that this is just of the
right order needed.

Let us choose a definite set of numbers to check the self--consistency of
the scheme.  Let $m_{\tilde{q}} = 700~GeV$ and $m_{\tilde{g}} \simeq 400~GeV$, so
that $x \simeq 0.3$.  We also choose, for this example, $\delta_{b_L b_R} = 0.1$.
The shift in $b$--quark mass from the gluino graph alone is then $\Delta m_b =
0.7~GeV$, which is small.  Including the chargino diagram, this shift may even
be smaller.  Next consider $V_{cb} \simeq M_{23}/m_b$.  Setting it equal to
$0.034$ determines $\delta_{s_L b_L} \simeq 0.15$.  Setting the diagonal mass of
the strange quark arising from the gluino exchange equal to $0.03~ GeV$, we get
$\delta_{s_R b_R} \simeq 0.65$.  Next we set $|V_{ub}| = |V_{us} V_{cb} - M_{13}/m_b|$
equal to $0.006$.  With a relative negative sign between the two terms, we find
$\delta_{d_L b_L} \simeq 0.065$.  This choice will now determine
$V_{us} \simeq M_{12}/m_s$, since there are no more parameters.
We find it is given by $V_{us} \simeq (11~MeV/m_s)$,
where $m_s$ is the $full$ strange quark mass at $\mu \sim 175~GeV$.  This
is of the right size for $m_s(\mu \sim 175~GeV) \simeq 0.05~GeV$ 
(corresponds to $m_s(1~GeV) \simeq 0.133~GeV$).  This shows the
consistency of the scenario.

  So far in the flavor mixings of the squarks, we have not assumed
left-right symmetry. If we however demand
left right symmetry is preserved by the soft supersymmetry breaking terms,
we will have an additional  mixing term
$\tilde d_R-\tilde b_R$.  Such a mixing entry will also enable us to induce
acceptable $d$--quark mass.
The strength of various operators would be related
by parity invariance.  The new operators induced by this new term are:
\begin{eqnarray}
        \tilde Q_1:&&Q_1({\rm L\leftrightarrow R}),
        \nonumber \\
        \tilde Q_2: && Q_2({\rm L\leftrightarrow R}) ,
        \nonumber \\
        \tilde Q_3: &&Q_3({\rm L\leftrightarrow R}) , \nonumber \\
    Q_4 &=&\bar d^\alpha_Rs^\alpha_L\bar d^\beta_Ls^{\beta}_R,\nonumber \\
    \tilde Q_4: &&Q_4({\rm L\leftrightarrow R}),\nonumber \\
    Q_5 &=&\bar d^\alpha_Rs^\beta_L\bar d^\alpha_Ls^{\beta}_R,
    \nonumber \\
     \tilde Q_5: && Q_5({\rm L\leftrightarrow R}) .\nonumber \\
\end{eqnarray}

These new operators will introduce new terms in the Hamiltonian:
\begin{eqnarray}
        H_{\rm extra}&=&-\frac{\alpha_s^2}{216 m_{\tilde{q}}^2}\Biggl\{
        \left(\delta^d_{12}\right)^2_{RR}
        \left(  24\,\tilde{Q}_1\,x\,f_6(x) + 66\,\tilde{Q}_1\,
        \tilde{f}_6(x) \right)
        \nonumber \\
        &+& \left(\delta^d_{12}\right)_{LL}\left(\delta^d_{12}\right)_{RR}
        \left( 504\,Q_4\,x\,f_6(x) - 72\,Q_4\,\tilde{f}_6(x)
        \right. \nonumber \\
        &+&\left. 24\,Q_5\,x\,f_6(x) + 120\,Q_5\,\tilde{f}_6(x) \right)
        \nonumber \\
        &+&  \left(\delta^d_{12}\right)^2_{RL}
        \left(  204\,\tilde Q_2\,x\,f_6(x) - 36\,\tilde Q_3\,x\,f_6(x) \right)
        \nonumber \\
        &+&\left(\delta^d_{12}\right)_{LR}\left(\delta^d_{12}\right)_{RL}
        \left( - 132\,Q_4\,\tilde{f}_6(x) - 180\,Q_5\,\tilde{f}_6(x)
        \right)\Biggr\}~.
        \end{eqnarray}

 The LLRR diagram can appear without a LR mixing now, i.e., $\tilde d_L-\tilde b_L$
 and $\tilde b_L-\tilde s_L$ on one line of the box diagram and
 $\tilde d_R-\tilde b_R$
 and $\tilde b_R-\tilde s_R$ on the other line of the same diagram.
The upper bound
 on the $\left(\delta^d_{12}\right)_{LL}\left(\delta^d_{12}\right)_{RR}$ term is
 shown in Table 3.

\begin{center}

 \begin{tabular}{|c|c|}  \hline
 $x$ & $\sqrt{\left|\Re \left(\delta^d_{12}\right)_{LL}
 \left(\delta^d_{12}\right)_{RR}\right|}$ \\
 \hline
0.1& $4.0\times 10^{-3}$ \\
0.3&$4.3\times 10^{-3}$ \\
2.0&$6.6\times 10^{-3}$ \\
4.0&$8.5\times 10^{-3}$ \\
8.0&$1.1\times 10^{-2}$ \\
10&$1.3\times 10^{-2}$ \\
\hline\end{tabular}
\end{center}

Table 3: Upper limits on the products of $\delta$'s
obtained for an average squark mass
 $m_{\tilde{q}}=700~\mbox{GeV}$ and for different values of
 $x=m_{\tilde{g}}^2/m_{\tilde{q}}^2$. For other values of $m_{\tilde{q}}$,
 the limits can be obtained by multiplying the ones in the Table by
 $m_{\tilde{q}}(\mbox{GeV})/700$.

Before we discuss the consequences of Table 3, let us comment on
the generation of $d$--quark mass purely from mixing with $s$ and $b$.
Clearly, that requires the $\tilde{d}_R-\tilde{b}_R$ mixing term.
There are two contributions to the $d$ mass, one from mixing with
the $b$ and the other from mixing with the $s$.  The former has a
magnitude $\sim V_{ub}^2 m_b$, while the latter goes as $V_{us}^2 m_s$.
Numerically the contribution from $d-s$ mixing is more important
and of the right magnitude.  So we focus on it.  For this estimate
to hold, the $\overline{s}_L d_R$ entry in the down quark mass
matrix should have the same magnitude as the $\overline{d}_L s_R$
entry.  The lower limit on the mixing parameters obtained by demanding
that $M_{21}^d \ge \theta_C m_s$ are the same as in Table 2, second
column.  (One can simply replace $L \rightarrow R$ in Eq. (14)
to get $M_{21}$.
All the relevant loop functions are the same.)  Similarly,
the upper limits from FCNC on  $\Re(\delta_{21}^d)_{LR}$
are identical to the numbers listed in Table 1, column 2.  This
is because of the parity inavariance respected by the gluino
box diagram.  We see that there is broad agreement between the
two sets of numbers, implying that reasonable $d$--quark mass
can be induced by pure mixing.

   From Table 3 we find the upper limit on the mixing term
  $\sqrt{\delta_{d_Lb_L}\delta_{b_Ls_L}\delta_{d_Rb_R}\delta_{b_Rs_R}}$.
Assuming $\delta_{d_Lb_L}\delta_{b_Ls_L}=\delta_{d_Rb_R}\delta_{b_Rs_R}$,  we
can use the bounds shown in Table 3 to put upper limit on
$\delta_{d_Lb_L}\delta_{b_Ls_L}$.
On the other hand, Table 2 gives the lower limit
on
$\delta_{d_Lb_L}\delta_{b_Ls_L}\delta_{b_Lb_R}$. If $\delta_{b_Lb_R}$ is $\sim 1/10$,
the lower limit surpasses the
upper limit when multiplied by this factor of 10. Thus, for the choice of
parameters we have made, it is not possible to maintain left-right symmetry
in the soft SUSY breaking terms. This is not a disaster for the model since
there may be hidden sector effects which may be able to generate the
necessary left-right breaking effects in the soft SUSY breaking sector.
This is especially so if the scale of $SU(2)_R$ breaking is higher than
that of supersymmetry breaking.
Another possibility is to use a different choice of parameter space
where $\delta_{b_Lb_R}$ $\sim 1$ so that bounds in the Table 3 do not contradict
 the assumption of left-right symmetry. This choice, for example, would correspond to
 large $\tan\beta \sim 60$,  with
$\mu\sim 1~ TeV$ ($m_0=450$ GeV and $m_{\tilde{g}}=1.5$ TeV).  Such a large value of
$\mu$ will require
 some fine tuning to obtain the Z-mass consistent with radiative
electroweak symmetry breaking, but is perhaps not out of question.

Other FCNC processes, such as from the loop induced $dsZ$ vertex or $ds$-Higgs vertex,
satisfy all the phenomenological constraints, mainly due to the decoupling behavior
of supersymmetric theories.

\section{Quark mixings from off-diagonal A terms}

The elements $M_{12}$ and $M_{23}$ can also be generated from the off diagonal A
terms. Again let us take a minimal flavor mixing structure.  We assume that
$A_{23}$ gives rise to $\tilde s_L\tilde b_R$ mixing and
$A_{12}$ gives rise to $\tilde d_L \tilde s_R$ mixing.  If left--right
symmetry is preserved, we need two more $A$'s induce
$\tilde b_L\tilde s_R$ and $\tilde s_L \tilde d_R$ terms ($A_{21}$ and $A_{32}$
respectively).  We shall assume
$A_{23}=A_{32}~ {\rm and} A_{12}=A_{21}$ in this case. We also have a
$\tilde b_L \tilde b_R$ term induced by the $\mu\tan\beta$ term as before.

The FCNC bound on $A_{12}$ can be obtained from the $K^0\bar{K^0}$. The
diagram will look like the fig. 1 with one cross on each squark line,
where the
cross represents the A term. The functions needed to calculate this diagram are
given in the Ref. \cite{masiero}. We find, for $m_{\tilde q}$=600 GeV and
$m_{\tilde{g}}=1$ TeV, $A_{12}v_d/m_{\tilde q}^2$ has to be less than
$7\times10^{-3}$. For the same
parameter space, the lower bound on the $A_{12}v_d/m_{\tilde
q}^2$ is $1 \times 10^{-3}$ from the one loop $\overline{d}_Ls_R$
mass diagram involving gluino and squarks.
The squark line will now have the $\tilde d_L \tilde s_R$
mixing term arising from $A_{12}$ (see Fig. 1).
The function $f1_M$ is used for this calculation.

The $\overline{s}_Lb_R$ mixing diagram will be generated by gluino mediation and
in the squark
line the off--diagonal $A_{s_Lb_R}$ appears. For $m_{\tilde q}$=600 GeV and
$m_{\tilde{g}}=1$ TeV we find the bound on the $A_{23}v_d/m_{\tilde q}^2 \ge0.009$ from the
mass diagram. There is no significant upper bound on $A_{b_Ls_R}$ term from the
available data.
Hence we will have a viable solution with left right symmetry intact.

Unlike the case of bi-linear mass terms, there could be two other indirect
bounds for the case of tri-linear $A$ terms.  These are the color and charge
breaking bounds (CCB) and unbounded from below (UFB) bound \cite{casas}.  If these bounds
are violated, the true vacuum of the MSSM may be either unstable or it may
be color and charge breaking. These indirect
bounds are not on the same footing as the FCNC bounds which are very direct.
It is not clear whether the CCB and UFB bounds are absolute bounds, they may
be evaded in various early universe scenarios \cite{kls}, for example if our universe
is in a metastable vacuum and if the tunneling rate to the true vacuum
is much slower than the age of the universe.  They may also be evaded if
non--renormalizable terms in the potential are taken into account or
equivalently, if the theory gives in to a new theory at a higher energy scale.
For completeness we list these CCB and UFB bounds below.

The most stringent (for our purpose) CCB bound is given as \cite{casas}:
\begin{eqnarray}
|A^{(d)}_{ij}|\leq \lambda^2_{dk}(m^2_{d_{L_i}}+m^2_{d_{R_i}}+m^2_{d_{1}}), ~k=Max(i,j)
\end{eqnarray}
where $m_1^2=m_{H_2}^2+\mu^2$.
The UFB bound is given as \cite{casas}:
\begin{eqnarray}
|A^{(d)}_{ij}|\leq \lambda^2_{dk}(m^2_{d_{L_i}}+m^2_{d_{R_i}}+m^2_{d_{\nu_m}}),~ k=Max(i,j)~.
\end{eqnarray}
The UFB and CCB bound requires $A_{23}v_d/m_{\tilde q}^2 \le 0.009$.
Hence we have a viable $A_{23}$ mixing.
However $A_{12}$ term is a problem.  From the UFB and CCB bound we have
$A_{12}v_d/m_{\tilde q}^2 \le0.0002$, which is smaller than the lower limit
required for generating $V_{us}$.
The problem with this  $A_{12}$ term happens because
the UFB or the CCB bound on this term
 involves the $s$-quark coupling. The  CCB bound involves $m^2_{\nu_m}$
and the UFB bound involves
$m_1^2(=m_{H_1}^2+\mu^2)$. So if we make  $m^2_{\nu_m}$ or $m_{H_1}^2$ larger,
 it is possible to have a viable $A_{1,2}$ mixing. FCNC constraint
can be satisfied in both the cases, although with a heavy supersymmetric spectrum.

In the left-right model leading to MSSM at low energies, these
constraints need not apply for the following reason: If we are working
at the $SU(2)_L\times SU(2)_R\times U(1)_{B-L}\times SU(3)_c$ level, it has been
shown that below the scale $v_R$, in addition to the interactions of
MSSM, there is an additional interaction of the form $W_{extra}= f_{ij} e^c_i
e^c_j {\Delta^c}^{--}$ \cite{kuchi}. This gives new quartic terms to the potential
that invalidate the CCB bounds of MSSM in the lepton sector. If we
extend the gauge group to $SU(2)_L\times SU(2)_R\times SU(4)_c$, there
can be analogous interactions involving quarks which can help
avoid the CCB bounds in the quark sector. Also in this class of models,
above $10^{10}$ GeV a new theory takes over which can substantially alter
the vacuum structure.

Let us now point out that while we have conducted the discussion of the
sections V-VII within the framework of the SO(10) or SUSYLR type models,
they are applicable to the $SU(3)_H$ models without any modification.
The only point worth noting here is that the presence of the horizontal
symmetry at a high scale implies that the soft breaking terms must originate
from basic interactions that are $SU(3)_H$ invariant after
horizontal symmetry breaking. Such a situation can always be arranged by
including extra $SU(3)_H$ multiplets (such as an $H$--octet) which do
not couple to matter fields and giving them appropriate VEVs. We do not
elaborate on these points since they involve standard methods.

\section{CP violation}

The simplest way to introduce CP violation into this model is by assuming
that the flavor mixing parameters $\epsilon_{ij}$ in Eqs. (6)-(7) are complex.
In Tables 1 and 3 we listed the lower limits on the flavor mixing parameters
arising from the {\it real part} of the $K^0-\bar{K^0}$ amplitude.  There are
constraints from the imaginary part too, they are more stringent by about
a factor of 200 compared to the numbers in Tables 1 and 3.  This suggests
that if the relevant phases that enter into the gluino box diagram
(see Fig. 2) are of order
$3 \times 10^{-3}$, these gluino box diagrams themselves can explain the observed value
of $\epsilon$ in the $K$ meson system.  Note that if the phases of $\epsilon_{ij}$
are of order $3 \times 10^{-3}$, the KM phase will be too small to account
for $\epsilon_K$.  This has implications for the B system.

If CP violation occurs only in the mixing, it might be though that this
is a superweak model of CP violation.  But it is actually not.
The direct CP violating parameter $\epsilon'/\epsilon$, receives a
one-loop contribution from the gluino penguin graph.  This diagram has
been calculated in Ref. \cite{barr}, and that result can be directly applied to the
model we are discussing.
It was shown in Ref. \cite{barr} that $\epsilon'/\epsilon=$
( 1 to 3)$ \times 10^{-3}$ (modulo hadronic uncertainty).  The reason for
such a relatively large value has to do with the chiral structure associated
with the gluino penguin graph of these models.

Thus there is ``approximate CP invariance" \cite{nir} in this class of models.
This means that all the phases are small, of order $10^{-2}$,
including the phases of the gluino
mass, $\mu$ parameter and $A$ and $B$ terms.
This is a welcome result since it can explain the smallness of neutron and
electron electric dipole moments, which is otherwise a puzzle in supersymmetric
models.  With all phases being order $10^{-2}$,  EDM of electron and neutron become consistent
with experiment.  (We assume that the contribution to the neutron EDM from the
strong CP phase is rendered harmless either by the axion solution or
by the constraints of left-right symmetry\cite{rasin}.)  KM model of
CP violation is not operative here.  This is especially significant for CP
asymmetries in the $B$ meson system.  In the KM model, many of the asymmetries
are large, of order 10\% or larger.  In our scenario, all asymmetries in the
$B$ system are expected to be too small.  The reason being the smallness of
all CP violating parameters.  (Recall that the phase in the KM model is not
small.)  If at the B factory, CP asymmetries are measured to be large, that
could exclude this scenario.

\section{Conclusion}

To summarize, we have discussed how the partial Yukawa unification that
results in a large class of SO(10) and SUSY left-right models as well
as a class of horizontal symmetry models provides a new way to
understand the smallness of quark mixing angles. Generating the appropriate
flavor mixings requires a specific pattern for the soft breakings.
As is well known, there are strong constraints on the allowed pattern of
soft breakings from flavor changing neutral current effects.
We have shown by a detailed analysis how the pattern
needed for our purpose is
consistent with the constraints of FCNC effects.  They are also compatible with
constraints arising by demanding that the vacuum be bounded from below and that
it conserves color and electric charge.

One place where this class of models may be subjected to experimental scrutiny
is CP violating asymmetries in the $B$ meson system which will be measured at
the B factories.  Unlike the KM model, all CP asymmetries in the $B$ system
are predicted to be too small, by the requirement of approximate CP invariance.
These models predict neutron and electron EDM near the present experimental
limits.  A third crucial test of the model involves the branching ratios
of the Higgs into fermion pairs.  Since the mass relation
$m_s^0/m_b^0 = m_c^0/m_t^0$ is corrected by suppressing $m_b^0$
via the one loop gluino contribution in one scenario, the higgs branching ratio
$\Gamma (h \rightarrow b \overline{b})/\Gamma(h \rightarrow \tau^+ \tau^-)$
will differ significantly from its standard model value \cite{kolda}.
The same diagrams that contribute to off--diagonal
quark and lepton mixings can lead to flavor violating Higgs decays.  Decays of the
Higgs into lepton pairs $h \rightarrow l_i^+ l_j^-$ for $i \neq j$ would
provide dramatic signatures in support of this class of models.  To have
an observable event rate, however, one would have wait for a Higgs factory such as the
muon collider.  In contrast to Ref. \cite{polonsky}, in our scheme the diagonal
muon mass arises already at the tree level, so no large effect is expected in
the $g-2$ of the muon.

\section*{Acknowledgments}

The work of KSB is supported by funds from the Oklahoma State University.
 RNM is
supported by the National Science Foundation grant No. PHY-9802551.

\begin{figure}[htb]
\centerline{ \DESepsf(fig34bdm.epsf width 12 cm) }
\smallskip
\caption {Gluino-squark diagrams for the generation of down quark masses and CKM
angles. }
\vspace{1 cm}

\centerline{ \DESepsf(fig12bdm.epsf width 12 cm) }
\smallskip
\caption {SUSY contributions to the $K^0-\bar K^0$ process.}
\vspace{0 cm}
\end{figure}

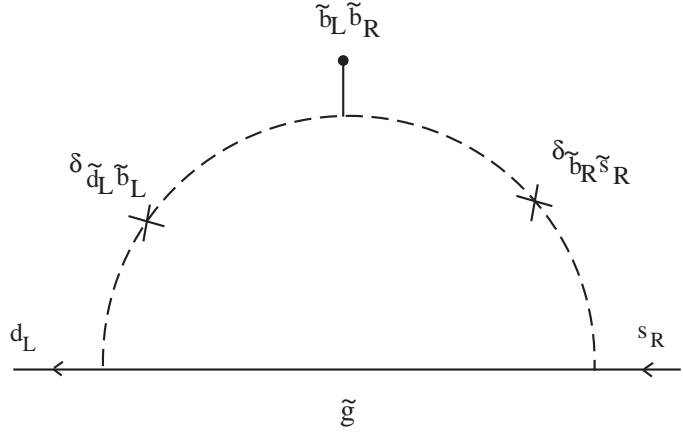
\end{document}